\documentclass[floats,aps,12pt]{revtex4}
\usepackage{epsf}

\usepackage[dvips]{color}

\begin{document}

\title{Actively Tuned and Spatially Trapped Polaritons}

\author{R.B. Balili$^{1}$, D.W. Snoke$^{1}$, L. Pfeiffer$^{2}$, and K.
West$^{2}$}
\affiliation{\mbox{$^{1}$} Department of Physics and Astronomy, University
of Pittsburgh, 3941 O'Hara St., Pittsburgh, PA 15260\\
\mbox{$^{2}$} Bell Labs, Lucent Technologies, 700 Mountain Ave., Murray
Hill, NJ 07974}


\begin{abstract}
{\bf Abstract}. We report active tuning of the polariton resonance of 
quantum well excitons in a
semiconductor microcavity using applied stress. Starting with the
quantum well exciton energy higher than the cavity photon mode, we use
stress to reduce the exciton energy and bring it into resonance with the
photon mode. At the point of zero detuning, line narrowing and strong increase
of the photoluminescence are seen. By the same means, we create an in-plane
harmonic potential for the polaritons, which allows trapping, 
potentially making
Bose-Einstein condensation of polaritons analogous to trapped atoms possible.
We demonstrate drift of the polaritons into this trap.
\end{abstract}

\maketitle

\newpage
  Microcavity polaritons have in the past decade been the object of great
interest for many 
scientists\cite{yama,deveaud,kav,dang,baum,ciuti,little,gia,science}
interested in the study of Bose-Einstein condensation (BEC). These 
particles, which are
mixed states of photons and excitons, have a very light mass, which in
principle allows them to condense at critical temperatures near room
temperature. For two-dimensional bosonic systems to truly condense at finite
temperatures, the application of potential traps or confinement in a 
region of finite
size is essential.\cite{mullin} Here we present a method to actively 
couple the exciton
mode to the cavity mode at fixed $k_{||} = 0$ and at the same time 
create an in-plane
spatial trap for both the lower and upper polaritons.
The polariton photoluminescence (PL) jumps up dramatically at 
resonance, and both the
PL and the reflectivity show line narrowing as the system approaches resonance.

The sample studied consists of three sets of four GaAs/AlAs quantum 
wells embedded
in a GaAs/AlGaAs microcavity, with each set of
quantum wells at an antinode of the confined mode, similar to the 
structure used in
previous work.\cite{yama} The cavity is designed in such a way that 
it is initially
negatively detuned, with
$\delta
\approx -20$ meV ($\delta = E_{\rm cav} - E_{\rm ex})$. A force is 
applied on back
side of the 150
$\mu$m thick substrate with a rounded-tip pin, with approximately 50 $\mu$m tip
radius, as shown in Figure 1. This pushes the exciton energy down 
toward the cavity
mode, at the same time creating a harmonic potential, following the 
method published
previously\cite{apl,ssc} for excitons. The harmonic potential is 
centered in the plane
of at the point of pin-sample contact.

\begin{figure}
\epsfxsize=.85\hsize
\epsfbox{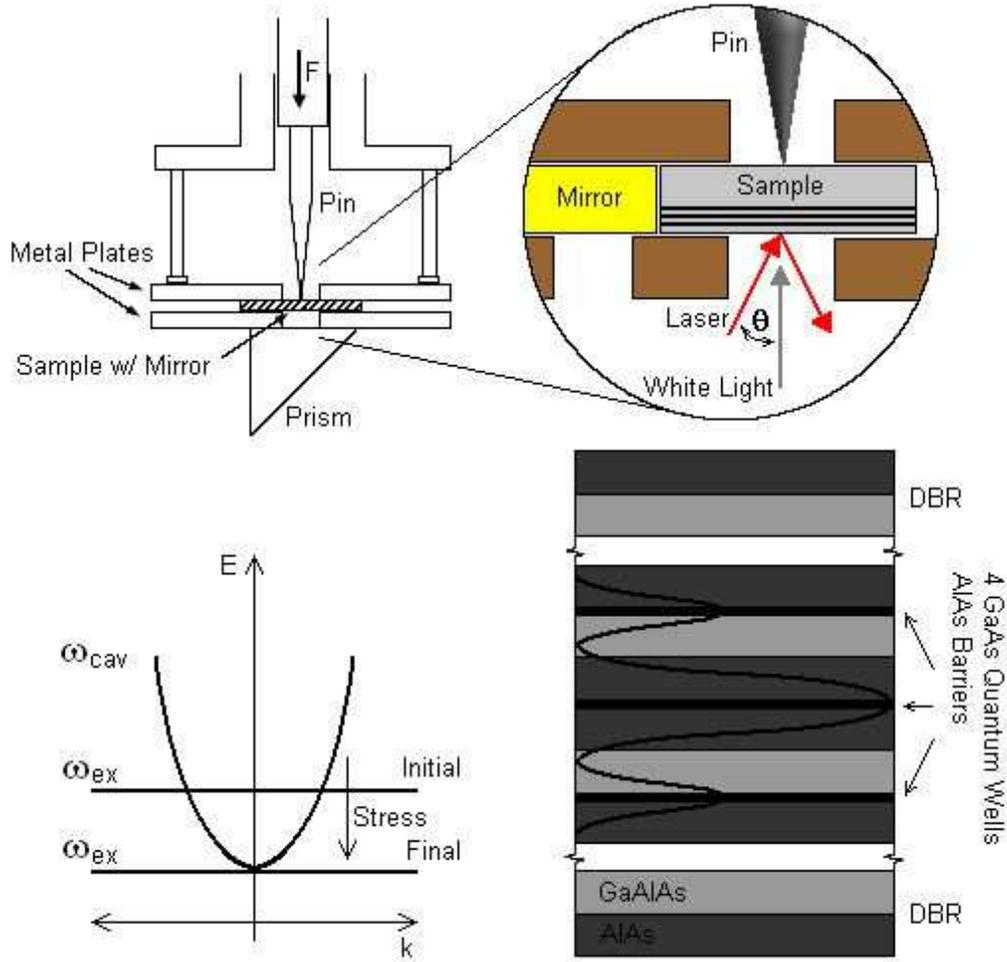}
\caption{Upper diagram: geometry of the experiment. Lower right: 
structure of the microcavity. Lower left: effect of stress tuning.
}
\end{figure}

Figure 2 shows the reflectivity spectrum as a function of position on 
the sample,
showing the anticrossing of the upper and lower polariton branches as 
the cavity
length is varied, due to the thinning of the layer thickness by about 
10\% toward the
edge of the wafer, which is part of the growth process. The pin 
stress point is chosen
several millimeters to the right of the crossover point, the point of
strongest coupling.  Figures 3 and 4 show photoluminescence and 
reflectivity data for
a sequence of increasing stresses applied to this sample. For the 
photoluminescence, a
helium-neon laser source (633 nm) is used to excite the sample 
off-resonantly, well
above the band gap, at $\theta = 12^{\circ}$ incidence, and defocused 
to a spot size of
several millimeters to cover the entire region of observation. 
Photoluminescence
emission collected normal to the sample is directed to a spectrometer 
and captured
with a Photometrics back-illuminated CCD camera.  For the sample reflectivity, a
collimated light beam (750 nm$-$1000 nm) is directed normal to the sample. The
reflected light is also collected normal to the sample. The mirror 
placed in the same
plane as the sample is used to normalize the sample reflectance. For all the
experiments, the sample was maintained at the temperature of 4.2 K. At this low
temperature, no luminescence is seen from the upper polariton. Upper polariton
emission for this sample starts to appear at about 40 K.

\begin{figure}
\hspace{.5cm}
\epsfxsize=.44\hsize
\epsfbox{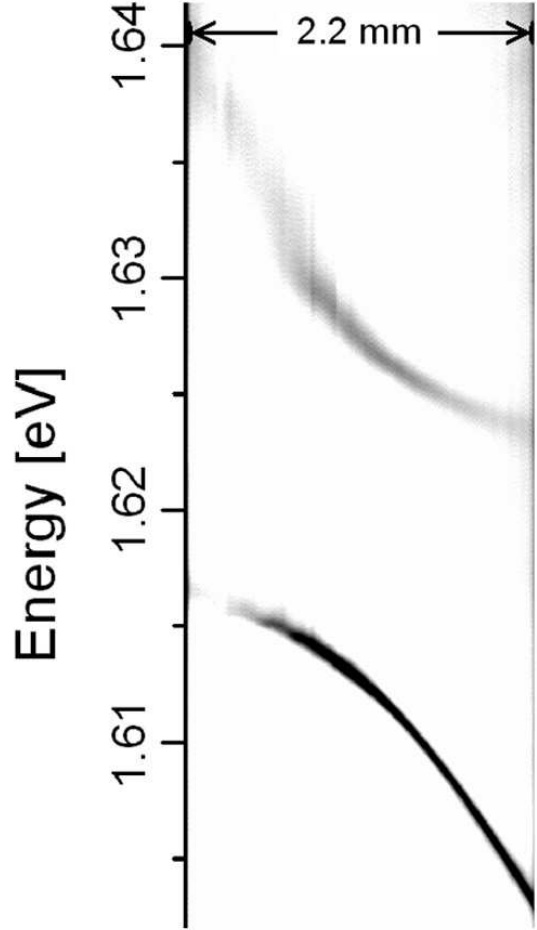}
\caption{Reflectivity spectrum as a function of position on the 
sample, for zero
stress. The polariton energy shifts as the cavity length shifts, due 
to the thinning
of the layers away from the center of the wafer.}
\end{figure}

As seen in these figures, a harmonic potential for both the upper and lower
polaritons is created. The polaritons are clearly in the strong 
coupling regime, since
if they were not, only the exciton states would respond to stress; 
the stress has
negligible effect on the dielectric constants of the materials and therefore
negligible effect on the cavity mode in the weak coupling limit. The energy gap
between the upper and lower polariton branches decreases, while the 
overall energy
shifts lower due to the band gap reduction.

\begin{figure}
\epsfxsize=0.7\hsize
\epsfbox{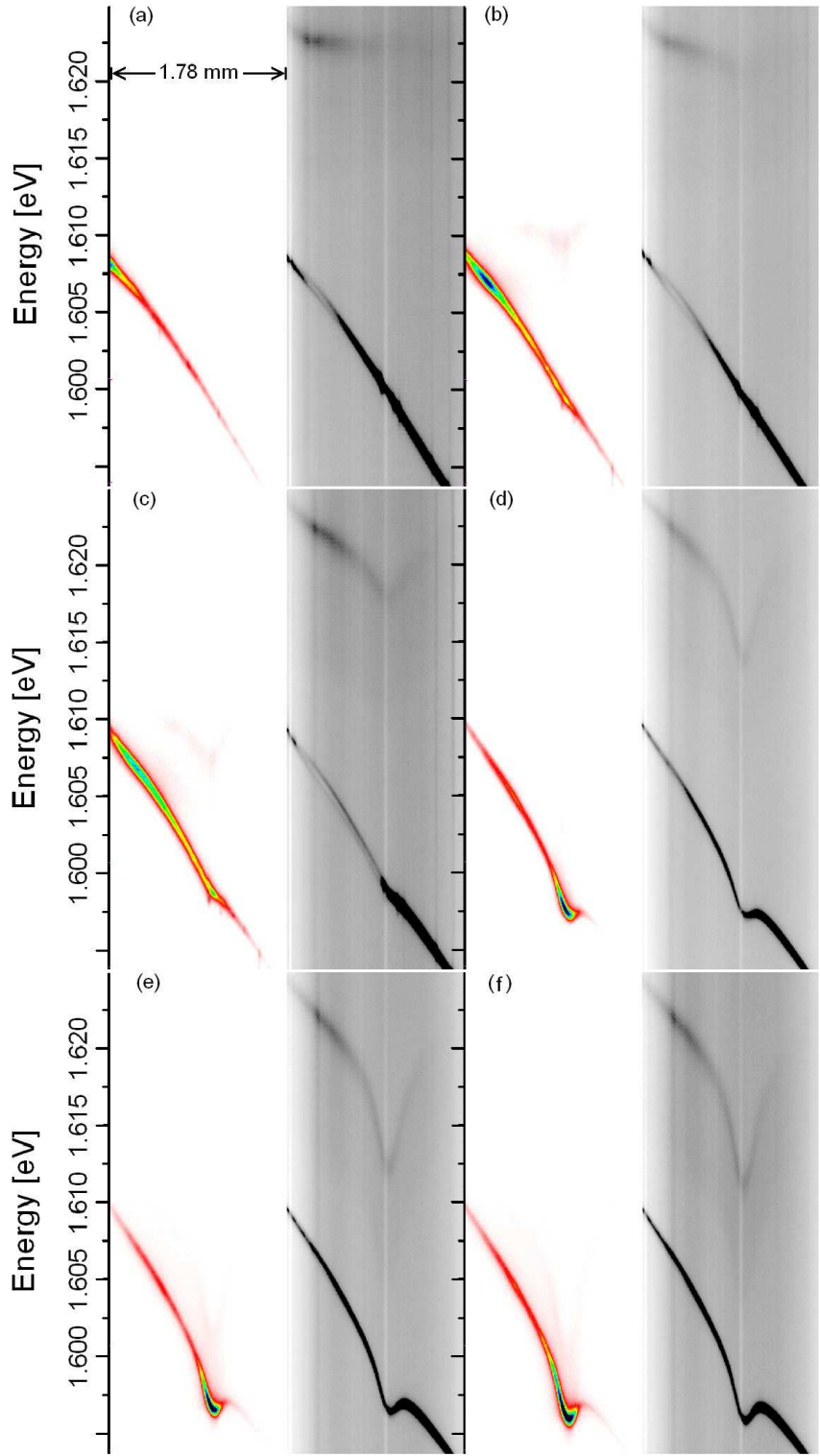}
\caption{
Left: Luminescence spectrum as a function of position on the sample, 
for various
levels of force on the pin stressor, (a) unstressed, (b) 0.75 N, (c) 
1.50 N, (d) 2.25
N, (e) 2.63 N, (f) 2.85 N (white: minimum; black: maximum intensity). 
These images
were created by illuminating the entire observed region (2.2 mm 
diameter) with a 5 mW
HeNe laser. Right: The corresponding reflectivity (black: 0.0; white: 
1.0). A harmonic
potential is clearly seen in both upper and lower polariton branches. }
\end{figure}

In addition to the energy shift of the bands, a striking increase of
the photoluminescence occurs, as seen in Figure 4. This is similar to 
the increase of
photoluminescence at resonance seen by tuning of the resonance
using a wedge of varying cavity thickness,\cite{stanley} but the 
increase in the
present case is dramatic, a factor of about 100.   The increase of the total
photoluminescence emitted from the front surface is consistent with 
an increase of
the coupling constant at resonance.   Consistent with the strong 
coupling, one can see
in Figure 4 the narrowing of the reflectivity spectra as the bare 
excitons and bare
photon modes approach resonance\cite{res} during stress tuning.

\begin{figure}
\epsfxsize=.9\hsize
\epsfbox{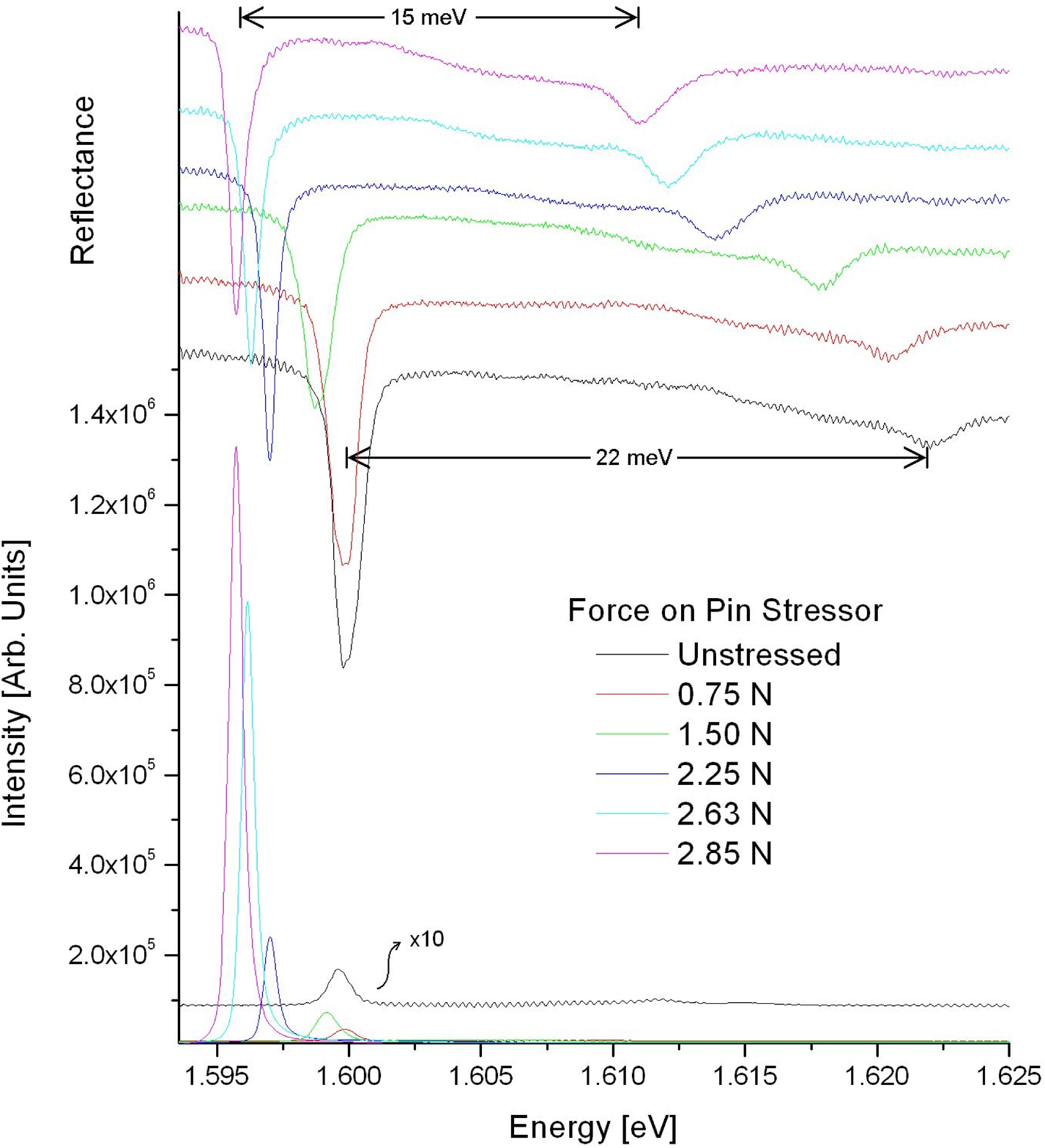}
\caption{
Top: Reflectivity at the bottom of the stress well, for a series of 
applied forces.
Bottom: Photoluminescence emission of the lower polariton, taken with the HeNe
excitation source (900 $\mu$W) focused (75 $\mu$m) at the bottom of the
stress well.}
\end{figure}

Since there is an energy gradient for the polaritons, one expects 
that they will
undergo drift.  Figure 5 shows spatially resolved
photoluminescence when the laser is tightly focused and moved to one 
side of the
potential minimum in the lower polariton branch. Drift is clearly 
observed over a
distance of more than 100 $\mu$m, similar to the drift seen 
earlier\cite{drift} for
polaritons in an energy gradient created by a wedge of the cavity thickness.
 The increase of the PL intensity 
seen in Figure
4 may be partly related to this effect, since polaritons will 
concentrate at the
bottom of the well instead of diffusing away from the excitation spot.

\begin{figure}[b]
\epsfxsize=.55\hsize
\epsfbox{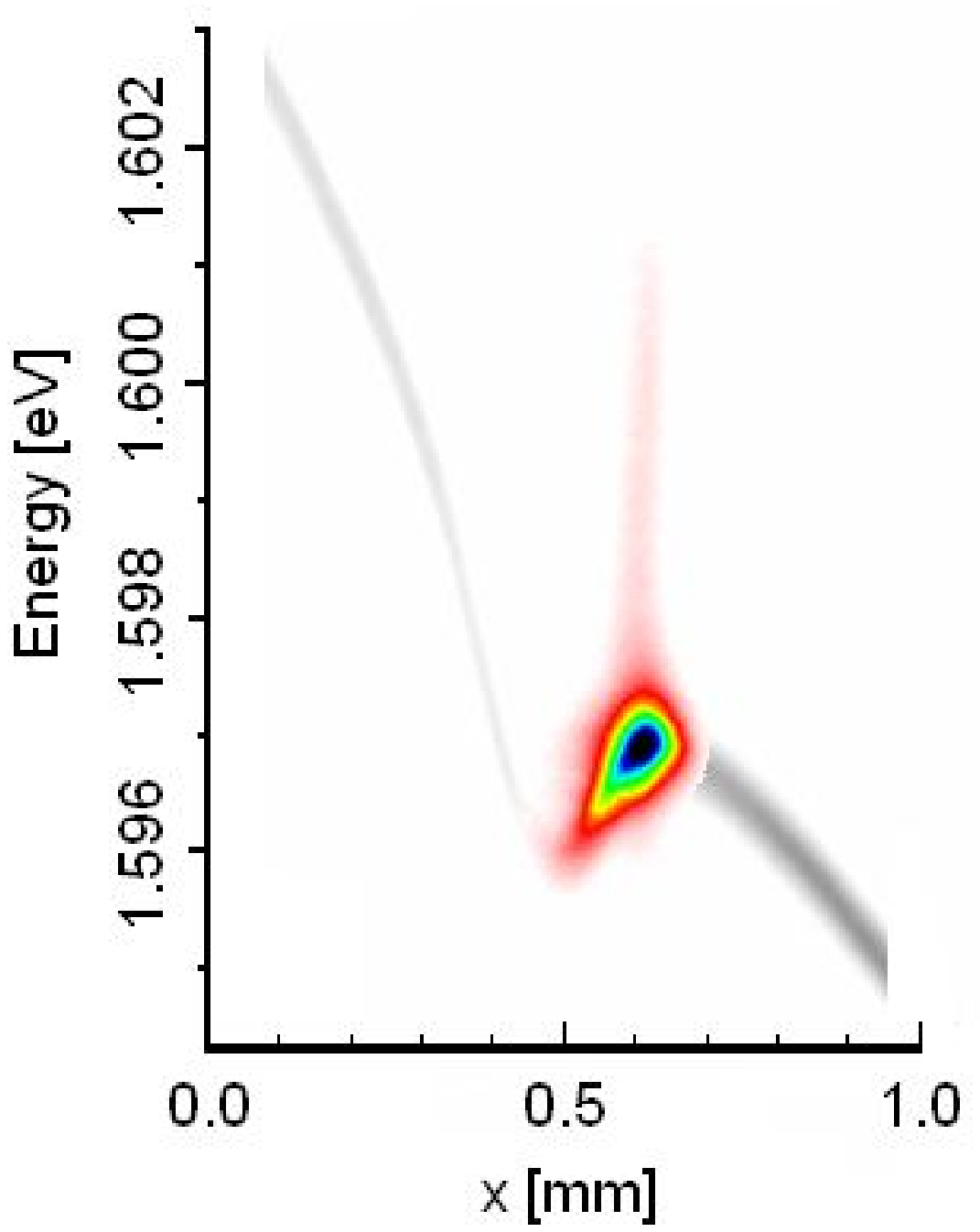}
\caption{
Spatially resolved photoluminescence for 2.85 N applied force (white: 
minimum, black:
maximum intensity) with the laser focused (75 $\mu$m) and shifted 
away from the bottom
of the stress well. The photoluminescence is superposed on the 
reflectivity spectrum
(gray) for the same conditions, to show the location of the well.}
\end{figure}

This method of trapping opens a wide variety of possibilities and promise
in the area of microcavity research and BEC of polaritons. 
Typically, only a tiny
region of a wafer is in the strong coupling regime, due to the wedge 
of the layer
thicknesses in standard growth processes. By using stress, one is no longer
limited to this small region; the method allows the freedom to use nearly any
part of the wafer and tune the bands to the region of strong 
coupling. Using electric
field to tune the resonance\cite{efield} has the drawback that the oscillator
strength of the exciton changes strongly with electric field. Also, 
as discussed
above, a harmonic potential minimum is essential for Bose-Einstein 
condensation of
polaritons or any other particles in two dimensions. The point of 
high stress becomes
a confining point for carriers, which can be used in a polariton 
laser. In previous
experiments,\cite{yama} the carriers were in free expansion with 
diffusion, with
energy shifts which depended on the local density.\cite{little} The present
experiments allow theory to treat a quasiequilibrium gas with a known confining
potential.

{\bf Acknowledgements}.
We wish to thank V. Hartwell, Z. V\"or\"os, and A. Heberle for the
invaluable comments and discussions, and H. Deng, G. Weihs, and Y. Yamamoto for 
contributions to
the design of this sample.  This material has been supported by the National
Science Foundation under Grant No. 0404912 and by  DARPA under Army 
Research Office
Contract No. W911NF-04-1-0075.


\end{document}